\documentclass[10pt,letterpaper]{article}
\usepackage[top=0.85in,left=2.75in,footskip=0.75in]{geometry}

\usepackage{changepage}

\usepackage[utf8]{inputenc}

\usepackage{textcomp,marvosym}

\usepackage{fixltx2e}

\usepackage{amsmath,amssymb}

\usepackage{cite}

\usepackage{nameref,hyperref}

\usepackage{microtype}
\DisableLigatures[f]{encoding = *, family = * }

\usepackage{rotating}

\raggedright
\usepackage[aboveskip=1pt,labelfont=bf,labelsep=period,justification=raggedright,singlelinecheck=off]{caption}

\makeatletter
\renewcommand{\@biblabel}[1]{\quad#1.}
\makeatother

\date{}

\usepackage{lastpage,fancyhdr,graphicx}
\usepackage{epstopdf}
\fancyheadoffset[L]{2.25in}
\fancyfootoffset[L]{2.25in}

\usepackage{color}

\newcommand{\reva}[1]{#1}
\newcommand{\revb}[1]{#1}
\newcommand{\revc}[1]{#1}

\begin{document}
\vspace*{0.35in}

\begin{flushleft}
{\Large
\textbf\newline{Topological phenotypes consitute a new dimension in the phenotypic space of leaf venation networks}
}
\newline
\\
Henrik Ronellenfitsch\textsuperscript{1},
Jana Lasser\textsuperscript{1},
Douglas C. Daly\textsuperscript{2},
Eleni Katifori\textsuperscript{3,1,*}
\\
\bigskip
\bf{1} Max Planck Institute for Dynamics and Self-Organization, 37077 G\"ottingen, Germany
\\
\bf{2} New York Botanical Garden, Bronx, NY, USA
\\
\bf{3} Department of Physics and Astronomy, University of Pennsylvania, Philadelphia, PA, USA
\\
\bigskip

%
%





* katifori@sas.upenn.edu

\end{flushleft}
\section*{Abstract}
The leaves of angiosperms contain highly complex venation
networks consisting of recursively nested, hierarchically
organized loops. 
We describe a new phenotypic trait of reticulate vascular networks based
on the topology of the nested loops. This phenotypic trait encodes
information orthogonal to widely used geometric phenotypic traits, and thus
constitutes a new dimension in the leaf venation phenotypic space.
We apply our metric to a database of 186 leaves and leaflets representing
137 species, predominantly from the Burseraceae family, 
revealing diverse topological network traits
even within this single family. We show that topological information
significantly improves identification of leaves from fragments
by calculating a ``leaf venation fingerprint'' from topology and geometry.
Further, we present a phenomenological model
suggesting that the topological traits can be explained by noise
effects unique to specimen during development of each leaf
which leave their imprint on the final network.
This work opens the path to new quantitative identification techniques 
for leaves which go beyond simple geometric traits such as vein density 
and is directly applicable to other planar or sub-planar 
networks such as blood vessels in the brain.

\section*{Author Summary}
Planar reticular networks are ubiquitous in nature and engineering,
formed for instance by the arterial vasculature in the mammalian neocortex,
urban street grids or the vascular network of plant leaves. 
We use a \reva{topological} metric to characterize the way loops are nested 
in such networks and analyze a large database of 186 leaves and leaflets,
revealing for the first time that the nesting of the networks'
cycles constitutes a distinct phenotypic trait orthogonal to previously
used geometric features.
Furthermore, we demonstrate that the
information contained in the leaf topology can significantly
improve specimen identification from fragments, and provide
an empirical growth model that can explain much of the observed data.
Our work can improve understanding of the functional significance of the
various leaf vein architectures and their correlation with the environment.
It can pave the way for similar analyses in diverse areas of research
involving reticulate networks.

\section*{Introduction}
The angiosperm leaf vein network fulfills the
combined requirements of efficient liquid transport
within the leaf and high robustness against load fluctuations and damage,
while at the same time providing structural reinforcement 
\cite{Katifori2010,Niklas1999,Sack2012,Sack2008}.

Modern leaf vein networks
evolved gradually from simple dendritic branching patterns by introduction
of anastomoses \cite{Sack2013, Roth-Nebelsick2001},
leading to leaf vascular networks that are highly reticulate, 
exhibiting nested, hierarchically organized vein loops.
The reticulate leaf vascular system is an example 
of evolutionary adaptation under various 
constraints \cite{Noblin2008,Jensen2013a,Jensen2013,Katifori2010,McCulloh2003}.

Despite some common trends, the diversity of vein morphology 
in dicotyledonous plants is striking (see for instance Fig.~1~a-f).
Current models of vascular development in the model species
\emph{Arabidopsis thaliana} predict several overlapping
phases in which the leaf primordium at first mainly grows by
cell division, then later by cell expansion \cite{Sack2012,Kang2004}. 
Lower order (major) veins are thought
to be formed during the first phases, whereas minor veins are formed
primarily during the later, leaving an imprint in the higher order
vascular system of the leaf.

The morphology, anatomy, and correlations with climate of the lower order 
\revc{vascular} 
architecture have been extensively studied \cite{Wright2004,Peppe2011}, and
primary and secondary vein traits can be easily quantified \cite{Sack2012}.
Certain leaf traits such as vein density are closely linked to 
photosynthetic efficiency 
\cite{Brodribb2007, Boyce2009, Brodribb2010}. \revc{Links to
climatic conditions and vegetation type have been proposed as well}
\cite{Peppe2011, Sack2013, Wright2005, Uhl1999}.

\revc{The hydraulic resistance of the whole plant is strongly affected by the
leaf hydraulic resistance. The smallest
veins, by virtue of their combined length and small
hydraulic diameter are responsible for the bulk of this resistance. At the same time, the smallest veins, and in particular
the small free-ending veinlets, are perhaps the most crucial for water delivery \cite{Fiorin2015}.}
However, the architecture of \revc{higher order vein reticulation} has been
largely ignored in the literature.
Other than an extensive descriptive nomenclature \cite{Ellis2009} and mainly 
qualitative measures \cite{Green2014}, 
to this day there is no quantitative work that goes beyond
obvious geometric characteristics, like minor vein density, areole size,
angle distribution, vascular segment length and width 
distribution \cite{Blonder2011,Sack2012,Bohn2002}. 
These characteristics by
themselves are not sufficient to describe the full architecture, in 
particular the organization of the loops. Loops typically show a large
degree of hierarchical nesting, i.e. larger loops composed of 
larger-diameter veins contain many smaller loops with smaller
vein diameter (see Fig.~1~e).

\reva{Although topological studies of spatial network architectures such as street networks are quite common
\cite{Barthelemy2011},
a detailed} quantitative characterization of \reva{the} topological properties 
related to reticulation has been elusive in the past, 
and only recently have researchers started
to seriously attack the question \cite{Katifori2012,Mileyko2012,Bohn2002}.

We use ideas inspired by computational topology \cite{Zomorodian2005} 
to define a 
metric suitable to quantify the architecture of higher order venation
of leaves. We apply our topological metric to a dataset of 186 leaves 
and leaflets,
demonstrating that our characterization constitutes a new phenotypic
trait in plant leaves and carries information \reva{complementary} to previously
used quantities. 
\reva{We then show that this information can be useful in the task
of identifying leaves from fragments, significantly improving
identification accuracy.}
We finally present a growth model that reproduces most of the
observed variation in the topological traits.

Our results suggest that topological and geometric venation traits
are \reva{approximately} independent, and that the higher order 
venation topology is mainly controlled by a small set of parameters regulating
noise during vein morphogenesis.

The topological venation traits we use can be employed in much
broader contexts than leaves, being applicable to any (sub-)planar,
anastomosing network such as blood vessels in the brain,
liver or retina, foraging networks build by slime molds, lowland river
networks, urban street networks or force chains in granular media,
thereby possibly opening up an entire new line of research.

\subsection*{Topological phenotypes}
Our topological metric quantifies the hierarchical nesting of loops 
within the network as 
well as the topological lengths of tapered veins.
The analysis follows an existing hierarchical decomposition algorithm  
\cite{Katifori2012, Mileyko2012, Bohn2005},
constructing from a weighted network a binary tree graph termed the
\emph{nesting tree} which contains
information about nesting of loops. The algorithm is schematically shown
in Fig.~1~g and discussed in the supplement. 


We stress that the method depends not on exact measurements of vein
diameters but only on relative order. Similarly, transformations which
slightly alter node positions do not affect the outcome (see 
Fig.~1~h). 

Once the binary nesting tree (see Fig.~1~g) has 
been obtained, its structure can be quantified. 
Here, for each node $j$ in  the nesting tree, we calculate
the nesting ratio $q_j = \frac{s_j}{r_j}$ \cite{VanPelt1992},
where $r_j \geq s_j$ are the numbers of leaf nodes in the right and left
subtrees of node $j$.
We then define the \emph{nesting number} as a weighted average
$i = \sum_j w_j q_j$, where $\sum_j w_j = 1$.
We employ an unweighted nesting number $i_u$, with
$w_j = 1$, and a degree-weighted nesting number $i_w$, 
with $w_j \propto d_j - 1 = r_j + s_j - 1$, where $d_j$ is called 
\emph{subtree degree}.
A high value of $i_{u,w}$ qualitatively represents graphs that are
highly nested such as those in the top row of Fig.~1~i.

The presence and extent of tapered veins is quantified as follows.
Starting from some edge $e$, we find the next
edge by taking the maximum width edge amongst all with smaller width
than $e$. We count how many steps can be taken until no more edges
with smaller width are adjacent, resulting in a topological
length $L_e$ assigned to each edge in the network. The mean topological
length $L_\mathrm{top} = \frac{1}{N_E} \sum_e L_e$, where $N_E$ is the 
number of edges, characterizes tapered veins in a network. 
Fig.~1~i shows a qualitative representation of
various example network topologies using mean topological length
and nesting number. 

Instead of using just the nesting \reva{number}, we additionally calculate 
pairwise topological
distances between networks as the two-sample Kolmogorov-Smirnov 
statistic $D_{KS}$ between the cumulative distributions of nesting
ratios in order to quantify the statistical similarity between nested
loop topologies.
\reva{Other methods to quantify the degree of topological dissimilarity between
binary trees representing biological systems have been proposed on the basis 
of a ``tree edit distance'' \cite{Ferraro2000}. Despite promise,
this distance suffers from being dominated by differences in the size
of the compared trees. In its local form \cite{Ferraro2004}, it suffers
from the opposite problem, quantifying only the similarity between 
the $n$ most similar subtrees.
In contrast, our method is designed to capture statistical
similarities between nesting trees, making it more suitable for dissimilarly
sized, noisy networks.}

\section*{Results}
We show that the topological characteristics described above provide 
a new dimension in the phenotypic space of leaf venation morphology.

For this, we analyze a dataset consisting of 186 leaflets from
various species primarily belonging to the
Burseraceae family (see \nameref{S1_Text} and \nameref{S1_Table}).  
Although most of species are 
therefore closely related, 
their venation patterns
show considerable diversity (see Fig.~1~a-f), rendering them a
good test set for our metrics 
The leaves were chemically cleared and stained to make their higher
order venation network apparent \cite{Vasco2014}, 
then scanned at high resolution ($6400\,\mathrm{dpi}$)
and vectorized in-house (see \nameref{S1_Text}).
Scanning whole leaves and digitizing at high resolution is 
computationally expensive but necessary
for this work to accurately represent the statistics of the high order 
veins \cite{Sack2014}. Publicly available
databases of scanned specimens \cite{Das2014}
contain mostly low resolution images.

\subsection*{Analysis of full leaf networks}
From the vectorized data, we obtained for each leaf five local geometric 
quantities: vein density $\sigma$ (total length of all veins/leaf area),
mean distance between veins $a$, mean areole area $A$, areole
density $\rho_A$, and average
vein diameter weighted by length of venation between junctions $d$.
The (un)weighted nesting number $i_{(u)\, w}$ was calculated from
all subtrees of the nesting tree with degree $d \leq 256$ in order
to remove leaf size effects for the full networks;
the mean topological length was calculated
from the whole network. 
Together, these metrics form a ``leaf venation fingerprint''
encompassing \emph{local} features of the network, that can be estimated from
leaf segments alone if necessary. 
\revc{Fig.~1~a shows the complete dataset plotted in the space of unweighted
nesting number and mean topological length. We plot the most abundant genera
\emph{Protium} (98 specimen in the dataset), \emph{Bursera} (21 specimen), 
and \emph{Parkia} (8 speciment) as different symbols. Although the dataset
does not allow for firm conclusions at this taxonomic level, both
\emph{Protium} and \emph{Parkia} appear to show a modest trend towards clustering 
around characteristic nesting numbers.}
We then employed Principal Component Analysis (see Fig.~2~b) and found that 
together, 
the first two principal components explain 81\% (=52\% + 29\%) 
of the total variance in the dataset.
Component 1 can be interpreted as containing mostly metrics derived from
geometry, whereas Component 2 contains mostly metrics from topology.
Topological lengths contribute roughly equally to either.
\reva{Even though small correlations between them exist, 
this} reveals local geometrical and topological leaf traits as 
\reva{approximately orthogonal traits} for the description of the phenotype of
leaf venation (see \nameref{S1_Text}, also for further analysis of the data
in terms of latent factors).

Pairs of leaves (see Fig.~2~a and Fig.~2~e,f) which
are close according to the topological distance defined by the
$D_\mathrm{KS}$ metric applied to the nesting ratio statistics can 
possess similar 
``by eye'' venation traits. In the samples in Fig.~2~e,f, cycle nestedness and vein thickness are traits that appear correlated. 
However, the topology of leaf venation constitutes a new phenotypic trait that provides information orthogonal to geometric traits.

\subsection*{Analysis of leaf fragments}
Topological information significantly helps in identifying leaf 
samples to species, especially when only a segment of the leaf is 
available. 
We fragmented all leaf samples in silico into
equally sized segments of ca. $1.2\times1.2$cm and calculated all 
venation traits for the individual pieces (see \nameref{S2_Table}). 
Here, we thresholded the nesting ratios at subtree degree $d \leq 128$.
We employed Linear Discriminant Analysis (LDA) \cite{Barber2012}
to classify the fragments based on specimen membership (see 
also \nameref{S1_Text}). 
We then calculated the the probability of correctly 
identifying a segment as belonging to one of the 186
leaves and leaflets (the accuracy, see Fig.~2~c).
Using only geometrical degrees of freedom, we found a 10-fold 
cross-validated accuracy of  \revc{$0.35$ (95\% CI: $[0.31, 0.39])$}. 
Adding topology improves the accuracy to 
\revc{$0.54$ (95\% CI: $[0.48, 0.60]$)}.
Additionally, for each pair of individual leaves in the dataset,
the same procedure was applied to obtain a mean pairwise accuracy
score (the probability of correctly identifying a fragment as belonging
to one of two leaves.) Again, using topological traits significantly 
improved the summary result (see Fig.~2~d and \nameref{S1_Text}). 
The same classification was applied towards identification of 
segments to species, as opposed to samples, with quantitatively 
similar results (see \nameref{S1_Text}).

\revc{It must be noted that there can be considerable variance among leaf
traits, even when comparing among specimen from a single plant ---
in particular between sun- and shade leaves \cite{Roth-Nebelsick2001,
Scoffoni2015}--- that should be taken into account if the information
is available.}

\subsection*{Comparison with venation growth model}
In order to explain the nesting ratio and topological length
distributions measured in our dataset, we
examine a developmental model for the formation of higher-order 
venation in which the interplay between strictly hierarchical loop genesis
and random noise is the major factor affecting nestedness.

\reva{Empirically, during the expansion growth phase of the leaf lamina,
high order vein loops grow and are subdivided by the appearance of new
veins, subsequent vein orders appearing
discretely one after the other \cite{Nelson1997,Kang2004}. Our model
intends to capture this phenomenological fact (see Fig.~3~a for an
illustration). The} model is compatible with models of
vein morphogenesis that invoke either auxin canalization \cite{Feugier2006}
or mechanical instabilities \cite{Laguna2008}, or a combination.
\revc{It is similar in spirit to that described in the 
    supporting information of \cite{Laguna2008} or \cite{Perna2011}
but adds fine-grained control of stochasticity.}

We stipulate that each leaf 
is subject to a species dependent characteristic amount
of noise during development, resulting in unique characteristic 
statistics of minor venation patterns.

The model as a whole is controlled by four dimensionless
parameters (see Methods section). 
In Fig.~3~b,c we show the distributions of normalized
areole size, mean topological lengths and nesting ratios for the same
two leaves as in Fig.~2~e,f. The real distributions can be
explained well by tuning two of the parameters.
Thus, noise during growth of cycles can explain the observed
local hierarchical nesting characteristics.

\reva{It should be noted that different mechanisms may underlie the
    organization of low order veins. Indeed, both models
    \cite{Fujita2006} and empirical observations \cite{Dengler2001}
    have found strong links between low order vein structure and leaf
    shape that may be connected to the overall growth pattern and developmental
    constraints of the lamina \cite{Couturier2009}.}

\section*{Discussion}
The leaf vasculature is a complex reticulate network, and properly chosen 
and defined topological metrics can quantify and highlight aspects of 
the architecture that have been ignored until now. The topological 
metrics presented in this work provide a new,
independent dimension in the phenotypic space of leaf venation, allowing
for more precise characterization of leaf features and improved
identification accuracy, including identification of fragments.
The extensive nomenclature for characterization of the vascular morphology 
\cite{Ellis2009} offers a discrete set of attributes that is 
mathematically insufficient to properly quantify a continuum of 
leaf venation phenotypes. However, this descriptive terminology can 
be incorporated as additional 
topological dimensions in the phenotypic space and alongside the metrics 
presented in this work can provide a tool to quantify inter- and 
intra- species diversity. 
In addition, we show that the local hierarchy of nested loops in
the leaf venation network can be explained by very simple stochastic
processes during development, pointing toward a universal 
mechanism governing (minor) vein morphogenesis.

The topological measures we employ have possible applications
that range far beyond the leaf data set explored here, being usable
on any loopy complex weighted network which possesses an embedding
on a surface. 
Examples of systems
that could benefit from an analysis along the lines of this work include
the blood vessels in the retina, liver or brain, anastomosing foraging 
networks built by slime molds and fungi, lowland river
networks, human-made street networks, force chain networks in granular
materials, and many more, thereby possibly opening up an entire
new line of research.

\section*{Materials and Methods}
\subsection*{Vectorization}
The extraction the networks from the original high-resolution scans
(6400 dpi) can be divided into two main steps: segmentation of the
image to create a suitable binary representation and skeletonization
of the shapes.
To segment the image we use a combination of Gaussian blurring to 
reduce noise, local histogram equalization and recombination with
the original image to increase contrast, and Otsu thresholding~\cite{otsu} 
to find the optimal threshold for the creation of the binary image.
For the skeletonization we use a vectorization technique known from 
optical sign recognition~\cite{vectorization1,zou-yan}. 
The approach relies on the extraction and approximation of the
foreground feature's contours using the Teh-Chin dominant point
detection algorithm~\cite{dominant_points} and subsequent triangulation 
of the contours via constrained Delaunay triangulation~\cite{CDT}. 
Therefore the foreground is partitioned into triangles which can be 
used to create a skeleton of the shape. Each triangle contributes a 
``center'' point to the skeleton which is determined by looking for
local maxima in the euclidean distance map~\cite{edm} of the binary
and together these center points approximate the skeleton. By looking
at edges shared between two triangles, neighborhood relations can be
established and an adjacency matrix can be created. This adjacency
matrix defines a graph composed of nodes (the former triangle centers)
and edges (the connections between two adjacent triangles). In 
addition to the topology of the graph the original geometry of the 
network including coordinates of the nodes and lengths and radii of
edges are preserved and stored in the graph.
The processing is done using algorithms implemented in \texttt{python}. 
The framework uniting all the aforementioned functionality is freely available at \cite{linkGithub}.

\revb{
\subsection*{Hierarchical decomposition}
A complete and detailed description of the hierarchical decomposition 
algorithm to extract the nesting tree from leaf network graphs can be found in
the supplement \nameref{S1_Text}. The software package used to
calculate nesting numbers, topological lengths, and geometric metrics
is freely available at \cite{linkHD}.
}

\subsection*{Modeling cycle nesting}
The model starts from a single rectangular
loop of veins (Fig.~3~a). The loops grow and subdivide when they reach a threshold size $A_0$ by introduction of a new vein. Not all loops subdivide at exactly the same size: the probability of subdivision as a function of areole area is a sigmoidal of width $\sigma_A$ (Fig.~3).  
All veins start with a fixed small width and grow linearly with time. The relative growth rate of
vein lengths and widths is controlled by the nondimensional parameter $\alpha$. The areole subdivision is only approximately symmetric: the new vein is randomly positioned close to the midline of the areole and the extent of the asymmetry is controlled by a parameter $\rho \in [0,1]$ (see 
\nameref{S1_Text}).
After the growing leaf has a certain size, the simulation is terminated and
random Gaussian noise with zero mean and standard deviation proportional to the parameter $f_n$ is added to the vein diameters.

The model is controlled by the four dimensionless
parameters $\rho$, $\beta=\sigma_A/A_0$, $\alpha$ and $f_n$.

\section*{Supporting Information}

\subsection*{S1 Text}
\label{S1_Text}
{\bf Detailed description of methods and further analysis.} 
Includes description of the geometric
and topological metrics used including more explicit hierarchical 
decomposition algorithm, an explanation of the leaf clearing,
staining and vectorization process, more details on the cycle growth
model. Further data analysis includes comparison of our data set with
earlier work, \revb{validation of the method}, and detailed results of 
PCA and Factor Analysis.

\subsection*{S1 Table}
\label{S1_Table}
{\bf Leaf fingerprint database.} The complete fingerprint data extracted
from the full leaf networks.

\subsection*{S2 Table}
\label{S2_Table}
{\bf Leaf fragment fingerprint database.} The complete fingerprint
data extracted from the $1.2 \times 1.2$ cm leaf fragments.



%
%
%


\newpage

\begin{figure}
    \includegraphics[width=\textwidth]{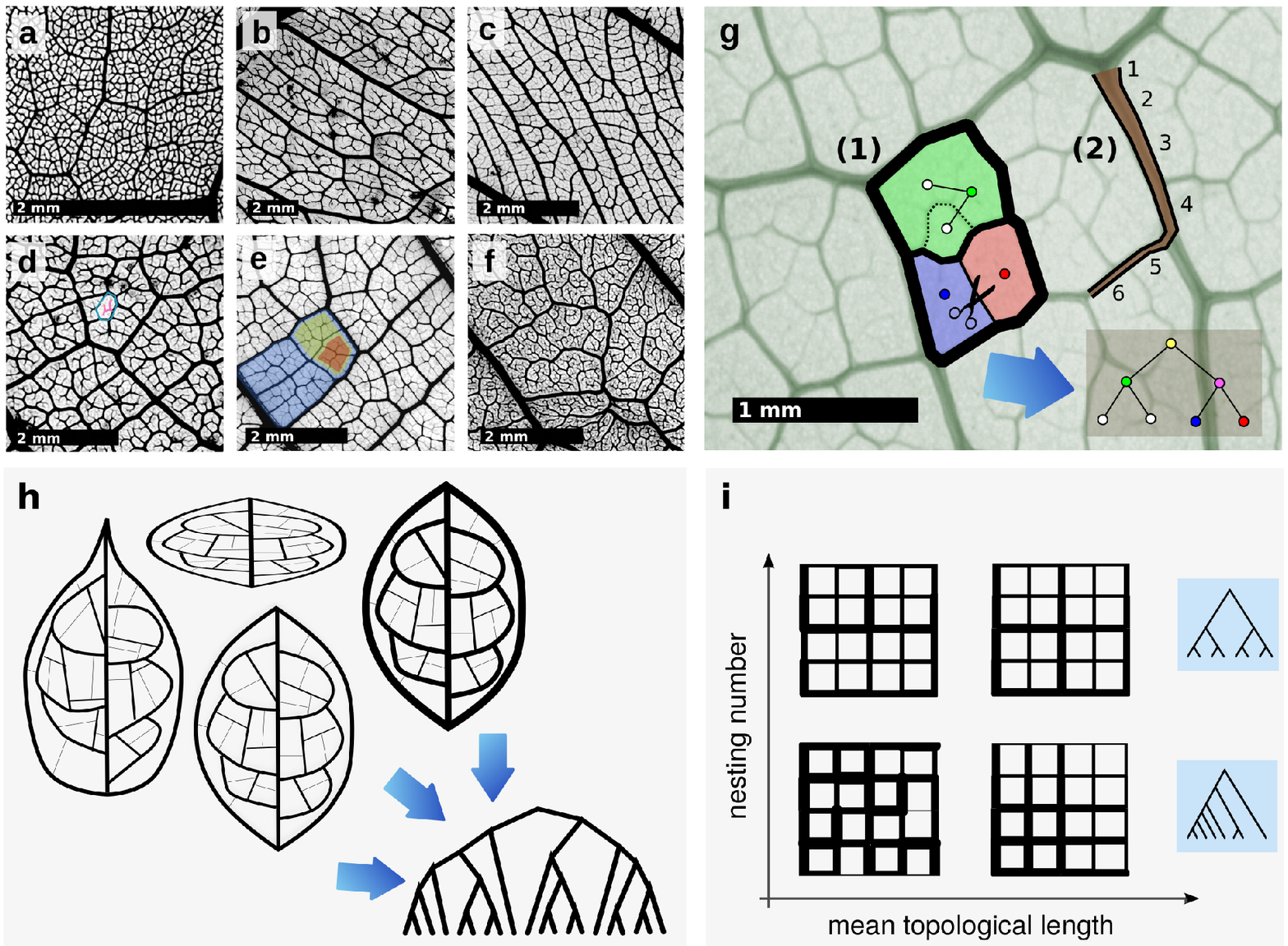}
    \caption{\textbf{a-f} Diversity of high order 
    leaf venation within the Burseraceae. \textbf{a} 
    \emph{Protium ovatum}.
    \textbf{b} \emph{Protium madagascariense}. \textbf{c}
    \emph{Pouteria filipes}. \textbf{d} \emph{Canarium betamponae}.
    A single areole is marked in blue, non-anastomosing highest order
    veins in red.
    \textbf{e} \emph{Brosimum guianensis}. The hierarchical nesting of
    loops is highlighted. 
    \textbf{f} \emph{Protium subserratum}. 
    \textbf{g} (1) The hierarchical decomposition algorithm.
    In a pruning step, all subgraphs which cannot be expressed as
    a superposition of cycles are removed. Then,
    the areoles are identified (facets) and assigned to leaf nodes in
    a tree graph (circles). The two facets whose intersection has
    minimum weight (vein thickness) are detected,
    their intersection is removed and their leaf nodes
    are connected to a new node which is identified
    with the new loop. The procedure is iterated until only
    one facet is left, which is identified with the root node.
    (2) Topological tapering length. Starting from a thick edge, we 
    walk on the network to that adjacent edge with the largest diameter
    that is still smaller than that of the current edge. This
    is repeated until no edge to proceed to can be found, and
    the number of edges traversed is counted.
    \textbf{h} Illustration of different leaf networks which result in
    the same hierarchical decomposition tree. Networks may undergo
    a geometric transformation such as stretching or squeezing. The
    vein thicknesses may change as long as their relative rank remains
    the same. Positions of junctions may change provided that no other
    junctions are crossed.
    \textbf{i} The space of topologies described by mean 
    topological length and nesting number. Shown are typical networks
    exhibiting various combinations of vein tapering and loop nestedness,
    as well as typical associated nesting trees.
    \label{fig:1}}
\end{figure}

\begin{figure}
    \centering
    \includegraphics[width=.6\textwidth]{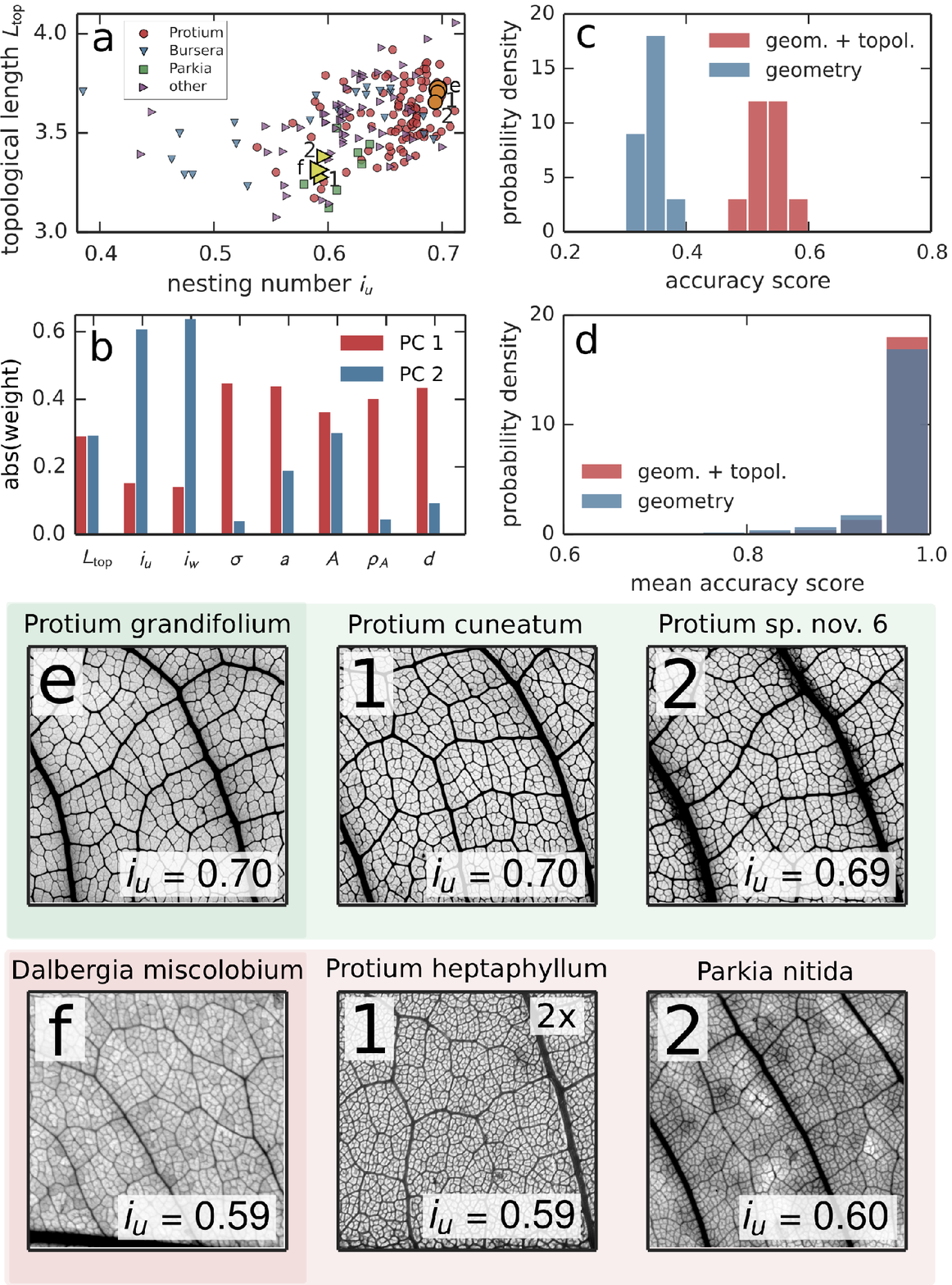} \\
    \caption{\textbf{a} Plot of the whole dataset consisting of 186
    leaf networks depending on the unweighted nesting number
    $i_u$ and mean topological length. One leaf of 
    \emph{Protium grandifolium} 
    (\emph{Dalbergia miscolobium})
    is marked by a circle (triangle) with black border as
    well as shown in e (f).
    The \revb{smaller} circles (triangles) in the same color show the two
    nearest neighbors according to the statistical distance $D_\mathrm{KS}$.
    \revc{The specimen belonging to the most abundant genera in the dataset are
        marked in order to assess predictive accuracy at a higher 
        taxonomic level. 
        The specimen belong to \emph{Protium} (98 specimen),
        \emph{Bursera} (21 specimen) and \emph{Parkia} (8 specimen).}
    \textbf{b} Weights of the 8 metrics, in the first two principal
    components of the dataset. Component 1 contains mostly geometry
    ($\sigma$, $a$, $A$, $\rho_A$, $d$),
    Component 2 mostly topology ($L_\mathrm{top}$, $i_u$, $i_w$), see also
    \nameref{S1_Text}.
    \textbf{c} Results of leaf identification from 
    fragments using Linear Discriminant Analysis (LDA). Accuracy scores 
    were obtained using 10-fold
    stratified cross-validation. The plot shows histograms
    of the resulting accuracy scores. Accuracy of identification
    is significantly improved when using both geometrical and topological
    information as opposed to only geometry.
    \revc{(Welch's $t(15.6)=15.8$, $p<0.001$)}.
    \textbf{d} Summary results of pairwise leaf identification from 
    fragments. All pairs of leaves were classified individually
    using LDA. Again, using topological traits significantly improves 
    the summary result (see \nameref{S1_Text}).
    \textbf{e, f} Images of the same
    leaves as those specially marked in \textbf{A} \revb{and their nesting
        numbers $i_\mathrm{u}$}, together with their
    nearest two neighbors 1, 2. 
    All images \revb{except for \textbf{f-1}} show a 
    $1\mathrm{cm} \times 1\mathrm{cm}$ gray-scaled \revb{and contrast-enhanced}
    crop of the original scan. \revb{Image \textbf{f-1} was zoomed in by
    a factor of 2 to show the nesting structure more clearly.} \label{fig:2}}
\end{figure}

\begin{figure}
    \includegraphics[width=\textwidth]{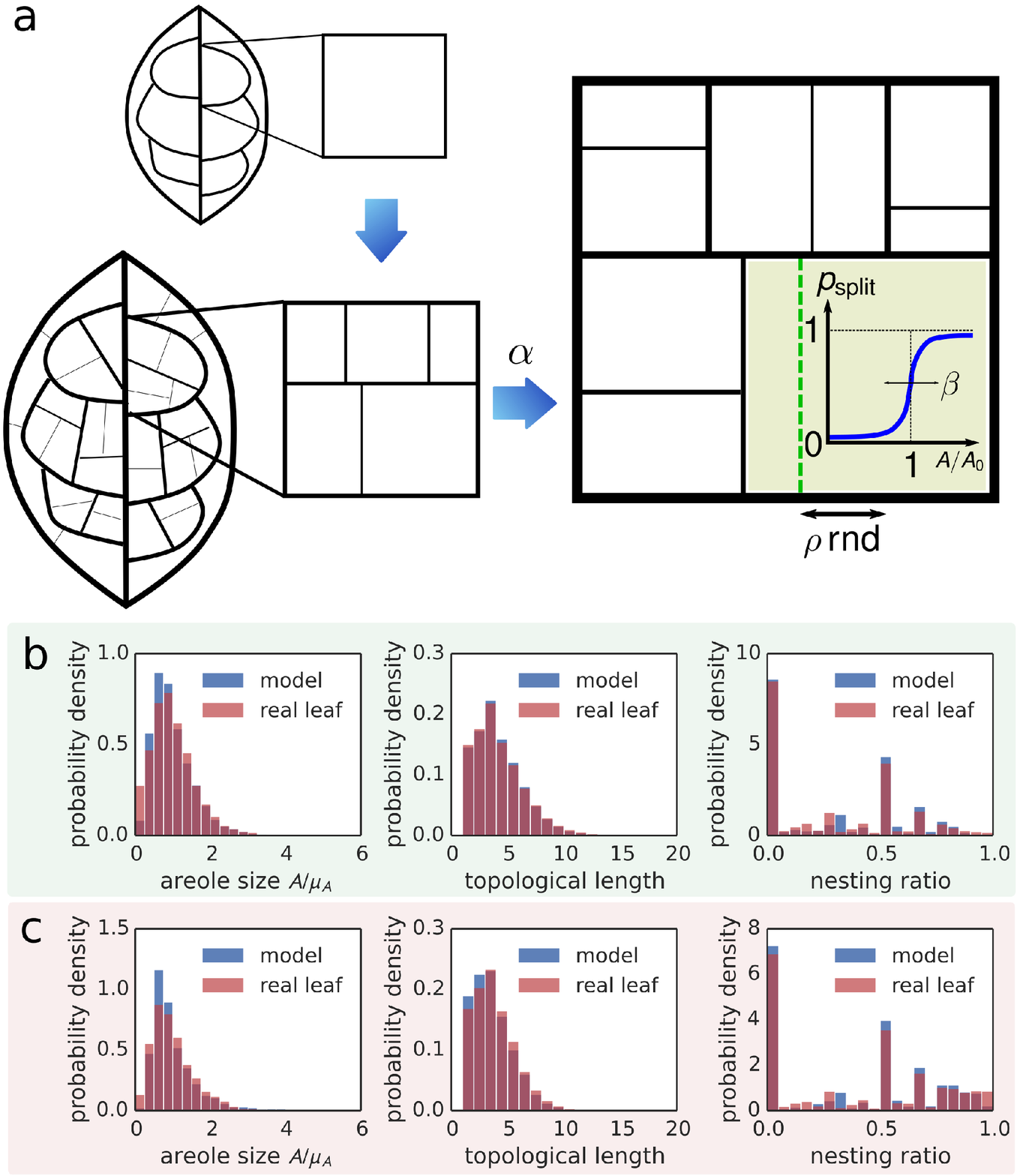}
    \caption{ \textbf{a} Developmental growth model.
    Beginning from a rectangular loop, new veins are introduced by
    successively splitting existing loops.  A new vein is formed at each
    iteration step with probability $p_\mathrm{split}$, which becomes
    significant as soon as loop area $A$ is close to a critical
    loop area $A_0$. $\rho\mathrm{rnd}$ determines the position of the new vein.
    \textbf{b} Comparison between developmental model and the same
    leaf as in Fig.~2~e, \emph{Protium grandifolium}. We compare the
    distributions of areole size $A$ normalized by the mean $\mu_A$,
    topological lengths and nesting ratios. The model parameters
    were $\alpha=0.25$, $\beta=0.5$, $\rho=0.2$, $f_n=0.1$,
    a low noise setting.
    The distributions agree well, explaining the strongly hierarchically
    nested structure in the high level venation network of 
    \emph{P. grandifolium}.
    \textbf{c} Comparison between developmental model and the same
    leaf as in Fig.~2~f, \emph{Dalbergia miscolobium}. 
    The model parameters
    were $\alpha=0.25$, $\beta=0.3$, $\rho=0.2$, $f_n=0.45$,
    a high noise setting. Again, the distributions agree reasonably
    well, explaining the relatively weakly nested high level venation
    network in \emph{D. miscolobium} by a large amount of noise
    in the vein widths.
    \label{fig:3}}
\end{figure}

\end{document}